\documentclass[prb,aps,twocolumn,amsmath,amssymb, showpacs, superscriptaddress]{revtex4}
\usepackage[pdftex]{graphicx}
 \usepackage{amsmath}
\usepackage[T1]{fontenc}
\usepackage{amssymb}
\usepackage{amsfonts}
\usepackage{bm}
 \usepackage{amsmath} 
\usepackage{lipsum}
\usepackage{amsfonts} 
\usepackage{amssymb, mathrsfs}
\usepackage{braket}
\usepackage{graphicx} 
\usepackage{subfigure}
\usepackage{bbm}
\def\beq{\begin{equation}}
\def\eeq{\end{equation}}
\def\bsp{\begin{split}}
\def\esp{\end{split}}
\def\bea{\begin{eqnarray}}
\def\eea{\end{eqnarray}}
\def\ba{\begin{array}}
\def\ea{\end{array}}

\def\dg{\dagger}
\def\sg{\sigma}
\def\up{\uparrow}
\def\dw{\downarrow}
\def\lb{\left(}
\def\rb{\right)}

\def\l.{\left.}
\def\r.{\right.}

\def\ra{\rangle}
\def\la{\langle}

\def\bo{\bold{k}}

\begin{document}

\title{Topological hard-core bosons on the honeycomb lattice}
\author{S. A. Owerre}
\email{solomon@aims.ac.za}
\affiliation{African Institute for Mathematical Sciences, 6 Melrose Road, Muizenberg, Cape Town 7945, South Africa.}
\affiliation{Perimeter Institute for Theoretical Physics, 31 Caroline St. N., Waterloo, Ontario N2L 2Y5, Canada.}

\begin{abstract}
This paper presents a connection between the topological properties of hardcore bosons and that of magnons in quantum spin magnets.  We utilize the Haldane-like hardcore bosons on the honeycomb lattice as an example. We show that this system maps to a spin-$1/2$ quantum XY model with a next-nearest-neighbour Dzyaloshinsky-Moriya   interaction.   We  obtain the magnon excitations of the quantum spin model and compute the  edge states, Berry curvature, thermal and spin Nernst conductivities. Due to the mapping from spin variables  to bosons, the hardcore bosons possess the same nontrivial topological properties as those in quantum spin system.  These results are important in the study of magnetic excitations in quantum magnets and they are also useful for understanding the control of ultracold bosonic quantum gases in honeycomb optical lattices, which is  experimentally accessible. 
\end{abstract}
\maketitle

  \section {Introduction}
In recent years, topologically nontrivial properties of band theory in electronic systems have become the driving force in condensed matter physics \cite{ yu6, yu7,  yu1, yu2, yu3, yu4, yu, yu5, yu8, fdm}.  Although electronic band theory has been studied over the past decades, the topological properties associated with it have only been appreciated recently. The Haldane model \cite{fdm} was the first example of quantum anomalous Hall effect (zero magnetic field quantum Hall effect) in electronic systems that originates completely from the topology of the energy bands. For many decades, this model was deemed as a toy model due to lack of experimental evidence. Quite recently, the experimental realization of Haldane model \cite{fdm} has been reported in ultracold fermionic atoms in a periodically modulated optical honeycomb lattice \cite{jot}. This observation has motivated numerous  theoretical studies of the effects of interactions in the so-called Haldane-Hubbard model \cite{pa,pa1,pa2,pa3}. 

On the other hand, topologically nontrivial energy bands occur frequently in ordered quantum spin magnets, however, their topological properties are rarely studied in quantum magnetism. The study of topological spin excitations (magnons) recently began with the work of Katsura-Nagaosa-Lee \cite{alex0} on the kagome and pyrochlore Heisenberg ferromagnets with  a nearest-neighbour (NN) Dzyaloshinsky-Moriya (DM) interaction \cite{dm, dm2}.  In addition to breaking the inversion symmetry of the lattice, the DM interaction plays the role of spin-orbit coupling as in electronic systems. As a result, the bosons acquire a phase (magnetic flux) while hopping on the lattice and  the system exhibits similar topological properties as in electronic systems. However, in bosonic systems there is no Fermi energy or Fermi level, and the idea of completely filled band does not apply. In this respect, the Chern number characterizing the nontrivial topology of the energy bands is, in fact, independent of the statistical nature of the particles. It merely predicts counter-propagating edge state modes in the vicinity of the bulk gap as a result of the bulk-edge correspondence. This leads to magnon edge state modes in ordered quantum magnets. Interestingly, the magnon edge state modes  carry a transverse heat (spin) current upon the application of a longitudinal temperature gradient. As magnons are uncharged particles, there is no Lorentz force,  the DM interaction plays the role of a magnetic field by altering the propagation of the magnon in the system, thus leads to { thermal Hall effect} dubbed {\it magnon Hall effect}. It was  discovered experimentally  by Onose  \textit {et al}., \cite{alex1} in the ferromagnetic insulator Lu$_2$V$_2$O$_7$ with a 3D pyrochlore lattice structure.  Recently, thermal Hall effect has been observed experimentally  on the 2D kagome magnet Cu(1-3, bdc) \cite{alex6}.  Besides, {\it phonon Hall effect} had been observed previously  \cite{stro,stro1, stro2, alex7a,stro3,stro4} in a completely different scenario. 

Mathematically, {\it magnon Hall effect} is manifested as a result of the nontrivial topology of magnon bulk band encoded in the Berry curvature $\boldsymbol{\Omega}(\bold k)={\nabla}_{\bf k}\times \bold{A}(\bold k)$, which acts a magnetic field, where $\bold{A}(\bold k)$ is a vector potential. This result was derived by Matsumoto and Murakami \cite{alex2} and relates the transverse thermal conductivity $\kappa_{xy}$ directly to the Berry curvature of the magnon bulk bands reminiscent of Hall conductivity in electronic  systems \cite{thou}.  It simply shows that the DM interaction can appear in any form provided the system exhibits a nontrivial topology in the magnon bulk bands. It was also shown that due to the boson population of the bands at high and low temperatures, thermal conductivity changes sign as function of temperature or magnetic field on the kagome and pyrochlore lattices \cite{alex4, alex5}. In addition, spin Nernst and torque effects have been recently proposed theoretically in these systems  \cite{alex7}.

The Heisenberg ferromagnet on the honeycomb lattice is known to exhibit Dirac points in the magnon energy bands \cite{jf}.  In this regard, we have proposed and studied a topological magnon insulator on the honeycomb lattice \cite{sol, sol1}.   We observed many distinctive features on the honeycomb lattice. Firstly,  the DM interaction appears as a next-nearest-neighbour (NNN) coupling transversing through the triangular plaquettes of the NNN sites on the honeycomb lattice as opposed to the kagome and pyrocholre lattices. It  generates a spin chirality on the triangular plaquettes of the NNN sites given by $\chi_A={\bf S}_i\cdot ({\bf S}_j\times {\bf S}_k)$ on sublattice $A$ and $\chi_B=-\chi_A$  on sublattice $B$, where ${\bf S}_i$ is the spin at site $i$. In the spin wave bosonic representation, the resulting Hamiltonian \cite{sol, sol1} is analogous to the Haldane model in electronic systems \cite{fdm},  which is known to possess a nontrivial insulating phase. Secondly,   thermal Hall conductivity $\kappa^{xy}$ does not change sign as a function of temperature or magnetic field and the low temperature dependence follows a $T^2$ law \cite{sol1}.  Recently,  spin Nernst effect has been studied in this system  \cite{kkim}. 

 In this paper, we show that the topological properties of magnons are not different from those of hardcore boson systems. This is due to a one-to-one correspondence between bosonic systems and quantum magnetic systems.  We study a Haldane-like hardcore bosons on the honeycomb lattice, which possesses a superfluid phase and a Mott phase. This model has a quantum Monte Carlo sign problem which hinders an explicit numerical simulation using this method. However, it can be studied by other available numerical schemes. We utilize the magnetic spin analogue of this model by mapping it to a quantum XY model with a DM interaction.  This enables us to study the magnetic excitations (magnons) of the quantum system.    We observe nontrivial topological properties of this model by studying the magnon energy bands, edge states, Berry curvature, thermal and spin Nernst conductivities of the corresponding quantum spin model of the hardcore bosons.  The correspondence between hardcore bosons and quantum spin systems  suggests that the associated hardcore boson model exhibits the same properties in the boson language,  because its excitations have to be magnons as well. These results are important as they can be studied in cold atom experiments.  In fact, the phase transitions between superfluid phase and Mott phase have been reported experimentally in   ultracold bosonic atom experiments  in optical lattices  \cite{mar, mar1, mar2, mar3, mar4, mar5}.  Hence, it is possible to realize magnon topological properties by using  cold atoms trapped in honeycomb optical lattices as has been reported in fermionic systems \cite{jot}.

\section{Haldane-hardcore bosons}
\begin{figure}[ht]
\centering
\includegraphics[width=2in]{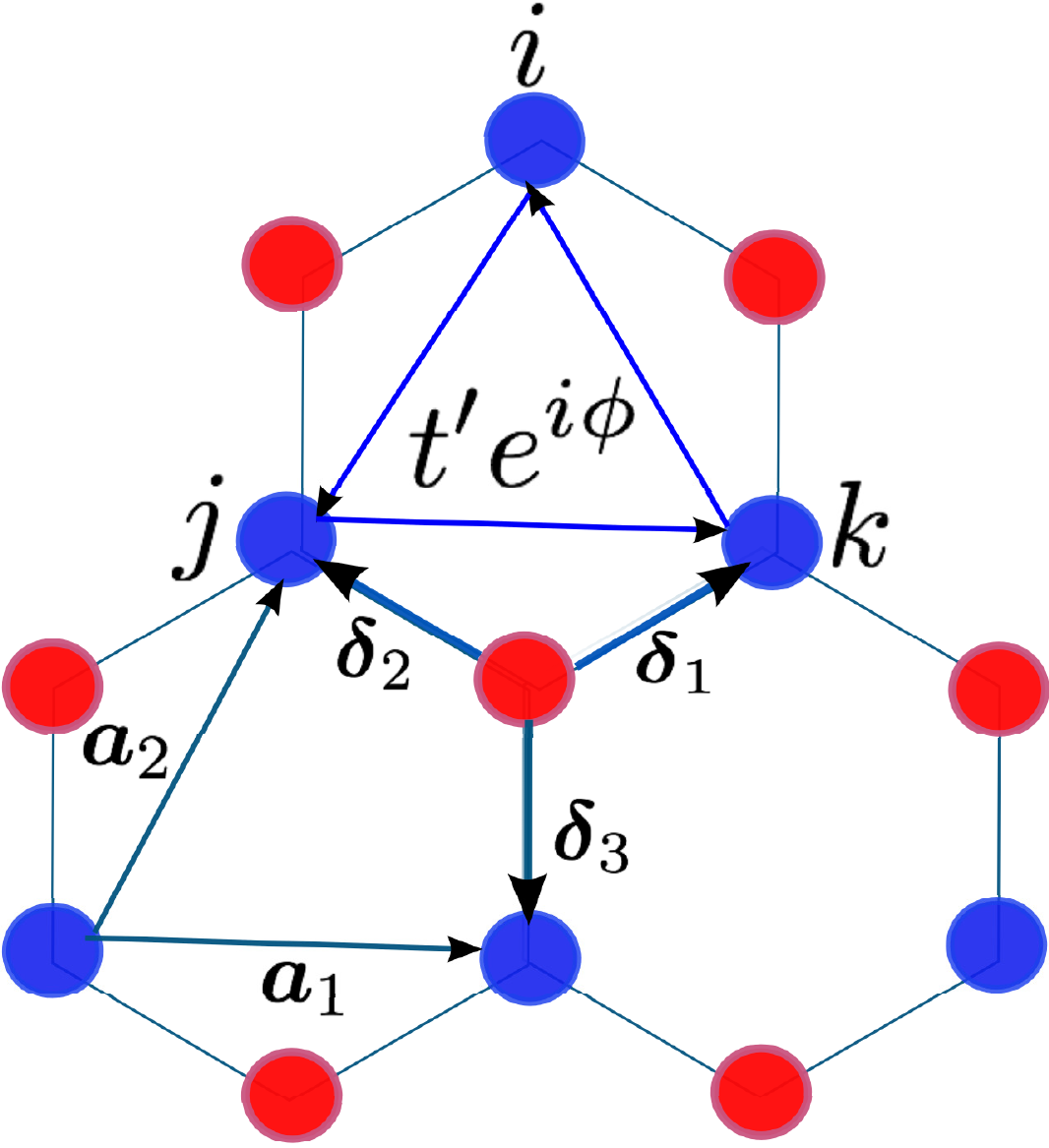}
\caption{Color online. The  honeycomb lattice with two sublattices indicated by different  colors.  The triangular arrows (labeled $i,j,k$) denote the direction of the  imaginary hopping amplitude, which corresponds to the  chiral DM interaction in spin variables. It generates a chirality on the triangular plaquettes. The coordinates are $\bold a_1=\sqrt{3}a\hat x;~ \bold a_{2,3}=a(-\sqrt{3}\hat x,\pm 3\hat y)/2$; $ \boldsymbol{\delta}_{1,2}=a(\pm\sqrt{3}\hat x,~\hat y)/2$, and $ \boldsymbol{\delta}_3=a(0, -\hat y)$. }
\label{unit}
\end{figure}

As mentioned above, the shortcoming of most topological bosonic models is that  quantum Monte Carlo (QMC) simulation is not applicable due to a deliberating sign problem. In this paper, we study one of these models in which QMC suffers a sign problem.   For hardcore bosons on the honeycomb lattice without any topological effects,  QMC has been utilized effectively in the study of the ground state phase diagrams \cite{alex8, alex8a}.  In this system there is a U(1) invariance,  hence the bosonic excitations (magnons) always exhibit a gapless (Goldstone) mode at the $\Gamma$ points ($\bold k =0$) and the two magnon bands exhibit Dirac points at the corners of the Brillouin zone $\bold K (\bold K^\prime)=(\pm 4\pi/3\sqrt{3}a, 0)$. Such  2D systems with Dirac nodes  have no topological effects in the excitation spectrum. The topological properties of these systems have received no attention mostly for the reason mentioned above. Fortunately,   the hardcore boson Hamiltonian  has a one-to-one correspondence to a spin-$1/2$ quantum XXZ or XY model \cite{matq}. Hence, the topological effects in this model can be studied in terms of the magnetic  excitations of the  quantum spin model.  As in many electronic systems, topological effects arise when a gap opens at $\bold K (\bold K^\prime)$. There are many different ways to open a gap at $\bold K (\bold K^\prime)$. The simplest way is to add a staggered on-site potential, as has been shown recently \cite{alex9, ban}. In the magnetic spin language,  this corresponds to a staggered magnetic field. An alternative way is to add a Haldane-like imaginary hopping interaction along the NNN sites, this would correspond to a DM interaction in the spin language  \cite{sol,sol1,kkim}. The resulting Hamiltonian of a gapped hardcore boson model on the honeycomb lattice can be written as

 \begin{align}
H&= -t\sum_{\langle ij\rangle}( b^\dg_i b_j +  h.c.) - t^\prime \sum_{\langle\langle ij\rangle\rangle}(e^{i\nu_{ij}\phi}b_i^\dg b_j +h.c.)\nonumber\\& -\mu\sum_i  n_i\label{hardcore},
\end{align}
where  $b^\dg_i$ and $ b_i$ are the bosonic creation and annihilation operators respectively;  $n_i=b^\dg_ib_i$; $\nu_{ij}=\pm 1$ depending on the hopping directions of the bosons as shown in Fig.~\ref{unit},  and $\phi$ is the phase of the hopping amplitude. The bosonic operators obey the algebra $[b_i, b_j^\dg]=0$ for $i\neq j$ and $\lbrace b_i, b_i^\dg \rbrace=1$.
Here,  $t>0$ denotes NN hopping, $\mu$ is the chemical potential, and $t^\prime$ is the NNN hopping with imaginary phase.  The model in \ref{hardcore}  has been recently studied in a different context \cite{pa1}. In that study, the authors did not investigate the mapping to quantum spin systems, but focused mainly on the Haldane-Hubbard Hamiltonian with a Hubbard on-site interacting potential. However, topological effects should also arise in noninteracting hardcore bosons \ref{hardcore}. In contrast to  Haldane model in fermionic systems \cite{fdm,jot},  the hardcore bosons \ref{hardcore}  can be mapped to a spin-$1/2$ quantum spin system \cite{matq}.   In the spin language, Eq.~\ref{hardcore} maps to 
 \begin{align}
H&=-J\sum_{\langle ij\rangle}(S_i^+S_j^-+ S_i^-S_j^+)+ \sum_{\la \la ij\ra\ra}{\bf D}_{ij} \cdot{\bf S}_{i}\times{\bf S}_{j}\label{hh1}\nonumber\\&-h\sum_i S_i^z,
\end{align}
where $S_i^{\pm}=S_i^x\pm iS_i^y$; ${\bf D}_{ij}=\nu_{ij}{\bf D}$, with ${\bf D}=2D\hat z$, and $\phi=\pi/2$. The new parameters are related to the hardcore boson variables by  $J\to t$, $\mu\to h$, $D\to t^\prime$.  Figure~\ref{unit} shows the direction of the fictitious magnetic flux generated by the  DM interaction, which corresponds to the imaginary hopping amplitude with  $\bold{D}_{ij}=\bold D$  for $i\to j$ on sublattice $A$ plus cyclic permutations,  and $\bold{D}_{ij}=-\bold D$ on sublattice $B$ for $i\to j$ etc.  As mentioned above, this alternating DM interaction has the effect on magnons as the chirality operators $\chi_A={\bf S}_i\cdot ({\bf S}_j\times {\bf S}_k)$ on sublattice $A$ and $\chi_B=-\chi_A$  on sublattice $B$, where ${\bf S}_i$ is the spin at site $i$.  It is important  to note that the spin Hamiltonian \ref{hh1} is unfrustrated in all the parameter regimes. It simply  describes  a ferromagnetic insulator.  Hence,  Eqs.~\ref{hardcore} and \ref{hh1} describe an ordered system and the magnon excitations of the spin system correspond to excitations of the hard-core bosons.
\begin{figure*}
  \subfigure[]{\includegraphics[width=.45\linewidth]{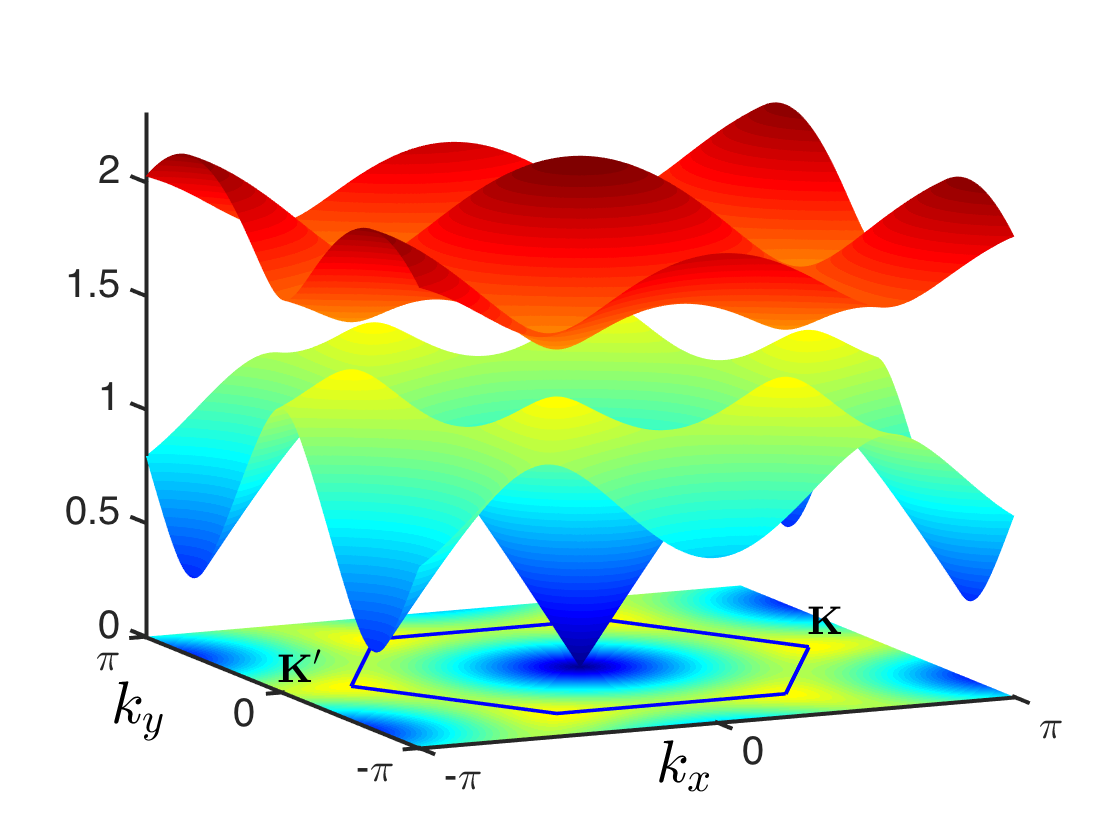}}
   \quad
   \subfigure[]{\includegraphics[width=.35\linewidth]{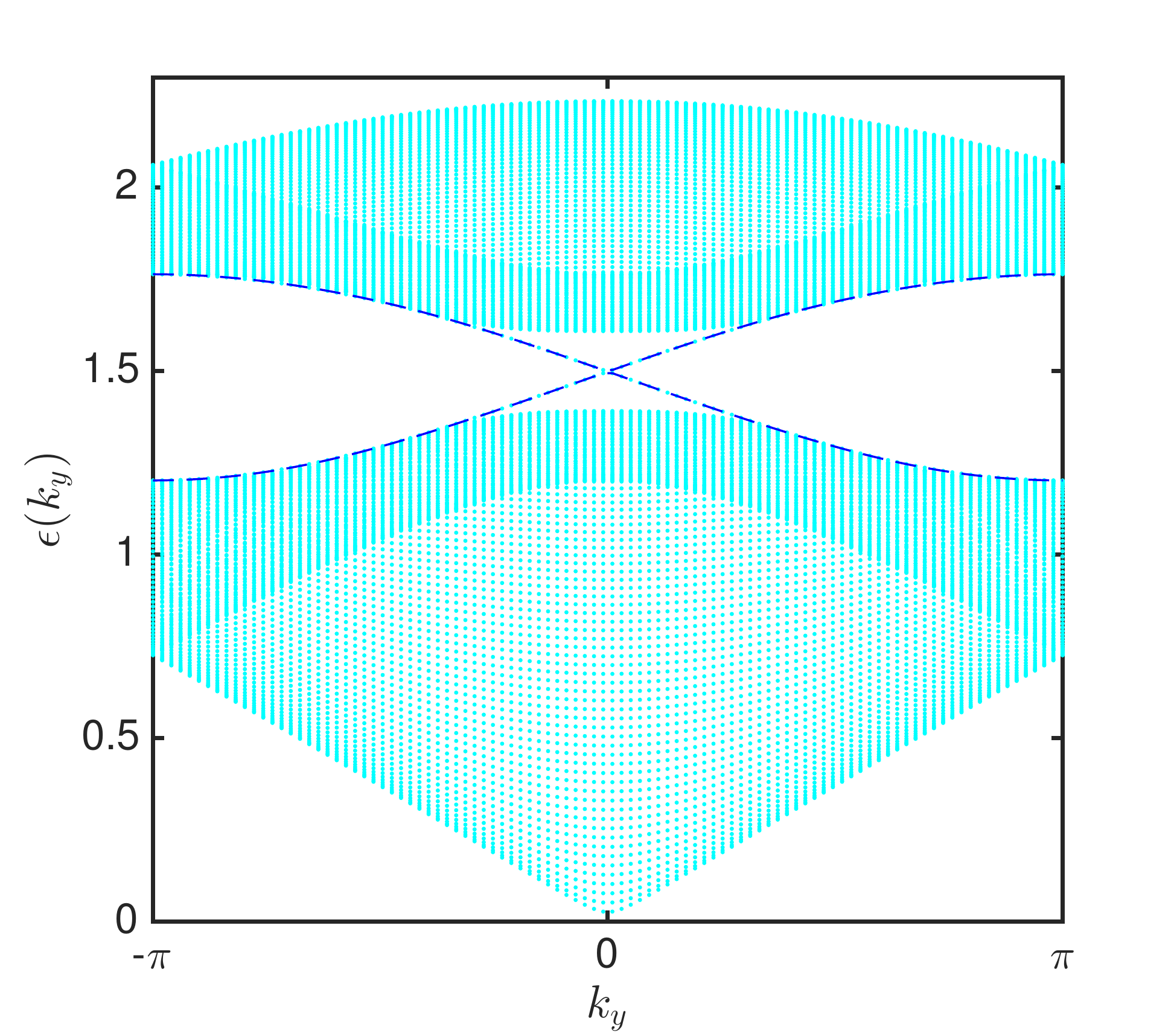}}
\caption{Color online. (a). The spin-$1/2$ magnon bulk bands of hardcore bosons in the chiral superfluid phase with $J=0.5;~D=0.25J$, $h=J$.  The blue hexagon is the Brillouin zone and the gap at $\bold K (\bold K^\prime)$ is $\Delta= 2|m(\bold k)|= 6\sqrt{3}v_D$. (b).   The spin-$1/2$ magnon bulk bands and the corresponding gapless edge states (dash lines) of one-dimensional armchair strip of honeycomb lattice with the same parameters.}
\label{band}
\end{figure*}
\section{Holstein-Primakoff formalism}
It is well-known that topologically nontrivial properties of magnon excitations  such as magnon edge states, thermal Hall effect, and spin Nernst effect, can be understood entirely  by  semiclassical approximations \cite{alex1, alex0, alex2,alex5,alex4, sol1, sol, kkim}.    Spin wave theory provides the simplest way to study the topological effects in ordered quantum magnetic systems.  In this section, we apply this formalism to the present model.
\subsection{Mean-field phase diagram.}
The mean field phase diagram provides the possible phases in the system.  In the mean field approximation, the spins can be approximated as classical vectors parameterized by a unit vector: $\bold{S}_i=S\lb\sin\theta_i\cos\phi_i, \sin\theta_i\sin\phi_i,\cos\theta_i \rb$. There are two sublattices on the honeycomb lattice as shown in Fig.~\ref{unit}.  The classical energy  with $\phi_i=0$ (coplanar spins) is given by
\begin{eqnarray}
E_c/NS &= -h_c\sin\theta_A\sin\theta_{B}-h(\cos\theta_A +\cos\theta_B), 
\label{cla}
\end{eqnarray}
where $N$ is the number of unit cells, $h_c=2JSz$ is the critical field, and $z=3$ is the coordination number of the lattice. At the mean-field level, the DM interaction does not contribute to the classical energy. The filling factor is given by $\rho= 1/2 + S(\cos\theta_A +\cos\theta_B)/2$. In this model, there are only two  possible phases --- a canted ferromagnet or superfluid at small magnetic field and a fully polarized ferromagnet or Mott phase, with the spins eqnarrayed along the $z$-axis at large magnetic field. In both ferromagnets $\theta_A=\theta_B=\theta$. The filling factor for the Mott phases are $\rho=0$ (empty) and $\rho=1$ (full) for $S_z=\mp S$ respectively. The phase boundary is  obtained by minimizing Eq.~\ref{cla} yielding $\cos\theta=h/h_c$.

\subsection{Magnetic excitations}
In ordered quantum magnets, it is customary to study magnetic excitations by linear spin-wave theory.    In the spin wave expansion, we first rotate the coordinate axes so that the $z$-axis coincides with the local direction of the classical polarization. 
\begin{align}
&S_i^x=S_i^{\prime x}\cos\theta  +  S_i^{\prime z}\sin\theta,\label{trans}~~
S_i^y=S_i^{\prime y},\nonumber\\&
S_i^z=- S_i^{\prime x}\sin\theta + S_i^{\prime z}\cos\theta.
\end{align}
Then, we express the operators in terms of the linearized Holstein Primakoff (HP) transformation, $S_{i}^{\prime z}= S-c_{i}^\dagger c_{i},~
 S_{i}^{\prime y}=  i\sqrt{ S/2}(c_{i}^\dagger -c_{i}),~
 S_{i}^{ \prime x}=  \sqrt{S/2}(c_{i}^\dagger +c_{i})$. The bosonic spin wave Hamiltonian becomes
 \begin{align}
H&=-\sum_{\la ij\ra}[v_{1}( c_{i}^\dagger c_{j}+ h.c.) +v_{2}( c_{i}^\dagger c_{j}^\dagger+ h.c.)]\nonumber\\&- v_D\sum_{\la \la ij\ra\ra}(e^{i\nu_{ij}\phi} c^\dagger_i c_{j}+h.c.)+v_0\sum_i c_{i}^\dagger c_{i},
\label{hp3}
\end{align}
where $v_{1,2}=JS(\cos^2\theta \pm 1)$, $v_D= 2DS\cos\theta$, $v_0=6JS\sin^2\theta+h\cos\theta$.  At zero magnetic field, the spins are aligned along the $x$-axis and $\theta=\pi/2$. Hence, the $z$-component of the DM interaction does not contribute in the linear spin wave expansion,  as it is not along the quantization axis.   In this case, one should take  ${\bf D}=D\hat x$.

As one can discern from Eq.~\ref{hp3},  the XY model does not have a simple analogue to Haldane model \cite{fdm} as the Heisenberg ferromagnets \cite{alex1, alex0, alex2,alex5,alex4, sol1, sol, kkim}. This is due to additional off-diagonal  term with  coefficient $v_2$. Fourier transforming Eq.~\ref{hp3} we obtain
\bea H= \frac{1}{2}\sum_{\bo}\Psi^\dg_\bo \cdot \mathcal{H}(\bo)\cdot\Psi_\bo + \text{const.},\eea 
where $\Psi^\dg_\bo= (\psi^\dg_\bo, \thinspace \psi_{-\bo} )$, and $\psi^\dg_\bo=(c_{\bo A}^{\dg}\thinspace c_{\bo B}^{\dg})$.

 The momentum  space Hamiltonian is given by 
 \begin{align}
\mathcal{H}(\bo)&= m(\bold k)\sigma_z\tau_z+ {\bold I}_\sg[{\bold I}_\tau v_0-\tilde v_{1}(\tau_+\gamma_\bo +h.c.)]\nonumber\\&- \tilde v_{2}\sigma_x(\tau_+\gamma_\bo +h.c.),
\label{hhhm}
\end{align}
where $m(\bold k)=v_D\rho_{\bold k}$ and  $\tilde v_{1,2}=zv_{1,2}$. We have introduced two Pauli matrices  $\boldsymbol\sigma$ and $\boldsymbol\tau$, where $\tau_\pm= (\tau_x\pm i\tau_y)/2$, while  ${\bold I}_\tau$ and ${\bold I}_\sg$ are identity $2\times 2$ matrices in each space.  The structure factors are  \bea 
\rho_{\bf k}=2\sum_\mu\sin {\bf k}\cdot\bold{a}_\mu;~\gamma_{\bf k}=\frac{1}{z}\sum_{\mu} e^{i{\bf k}\cdot \boldsymbol{\delta}_\mu};\eea  where  $ \boldsymbol{\delta}_\mu$ and $\boldsymbol{a}_\mu$ are the NN  and NNN vectors shown in Fig.~\ref{unit} respectively.

\subsection{Magnon bands and Berry curvatures}

The spin wave Hamiltonian in Eq.~\ref{hhhm} is Hermitian but it is not diagonal.  It is diagonalized by a matrix $\mathcal{U}(\bo)$ via the transformation $\Psi^\dg_\bo\to\mathcal{U}(\bo)\tilde{\Psi}^\dagger_\bo$, which satisfies the relation \bea \mathcal{U}^\dg \mathcal{H}(\bo) \mathcal{U}= \epsilon(\bo); \quad \mathcal{U}^\dg \eta \mathcal{U}= \eta,\eea  with $\eta=\sigma_z{\bold I}_\tau$. The matrix $\tilde{\Psi}^\dagger_\bo$ contains the Bogoliubov operators $(\alpha_{\bo}^{\dg},~\beta_{\bo}^{\dg})$ and  $\epsilon(\bo)=[\epsilon(\bo),-\epsilon(\bo)]$ are the  eigenvalues. This transformation leads to a non-Hermitian Bogoliubov Hamiltonian  $\mathcal{H}_B(\bo)=\eta\mathcal{H}(\bo)$, which  is given by
 \begin{align}
\mathcal{H}_B(\bo)&=\sigma_z{\boldsymbol a}(\bo)+i\sigma_y {\boldsymbol b}(\bo)+m(\bold k){\bold I}_\sg\tau_z,
\label{ham3}
\end{align}
where
\begin{align}
{\boldsymbol a}(\bo)&=\tau_0 v_0 - \tilde v_{1}(\tau_+\gamma_\bo +h.c.),\\{\boldsymbol b}(\bo)&=-\tilde v_{2}(\tau_+\gamma_\bo +h.c.).
\end{align}
The eigenvalues of the non-Hermitian Bogoliubov Hamiltonian give the magnon energy bands. 
 For $m(\bold k)=0$, the matrices  ${\boldsymbol a}(\bo)$ and ${\boldsymbol b}(\bo)$ are constants of motion and can be diagonalized separately. In this case  the Hamiltonian corresponds to the conventional XY model (or hardcore bosons) and the energy bands are gapless at  $\bold K (\bold K^\prime)$. For $m(\bold k)\neq 0$ this symmetry is broken and dynamics are introduced into the system. 
The positive eigenvalues are given by
\begin{align}
\epsilon_\lambda(\bo)=\sqrt{\lb v_0 + \lambda \sqrt{\lb \tilde v_{1}|\gamma_\bo|\rb^2+[m(\bold k)]^2}\rb^2-\lb \tilde v_{2}|\gamma_\bo|\rb^2},
\label{xym}
\end{align}
where $\lambda=\pm$ labels the top and the bottom bands of the positive energies.
\begin{figure}
\includegraphics[width=1.1\linewidth]{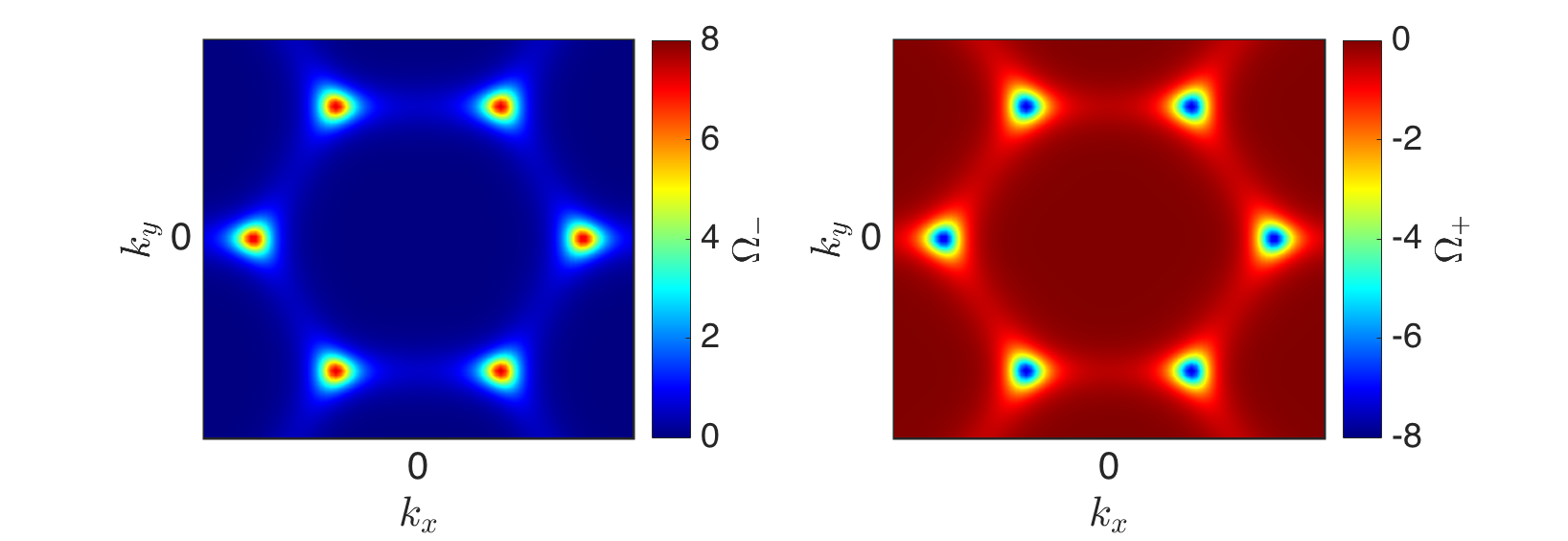}
\caption{Color online. The spin-$1/2$ magnon Berry curvatures of the hardcore bosons for the top and bottom bands of Fig.~\ref{band}.}
\label{berry}
\end{figure}
\begin{figure*}
  \subfigure[]{\includegraphics[width=.45\linewidth]{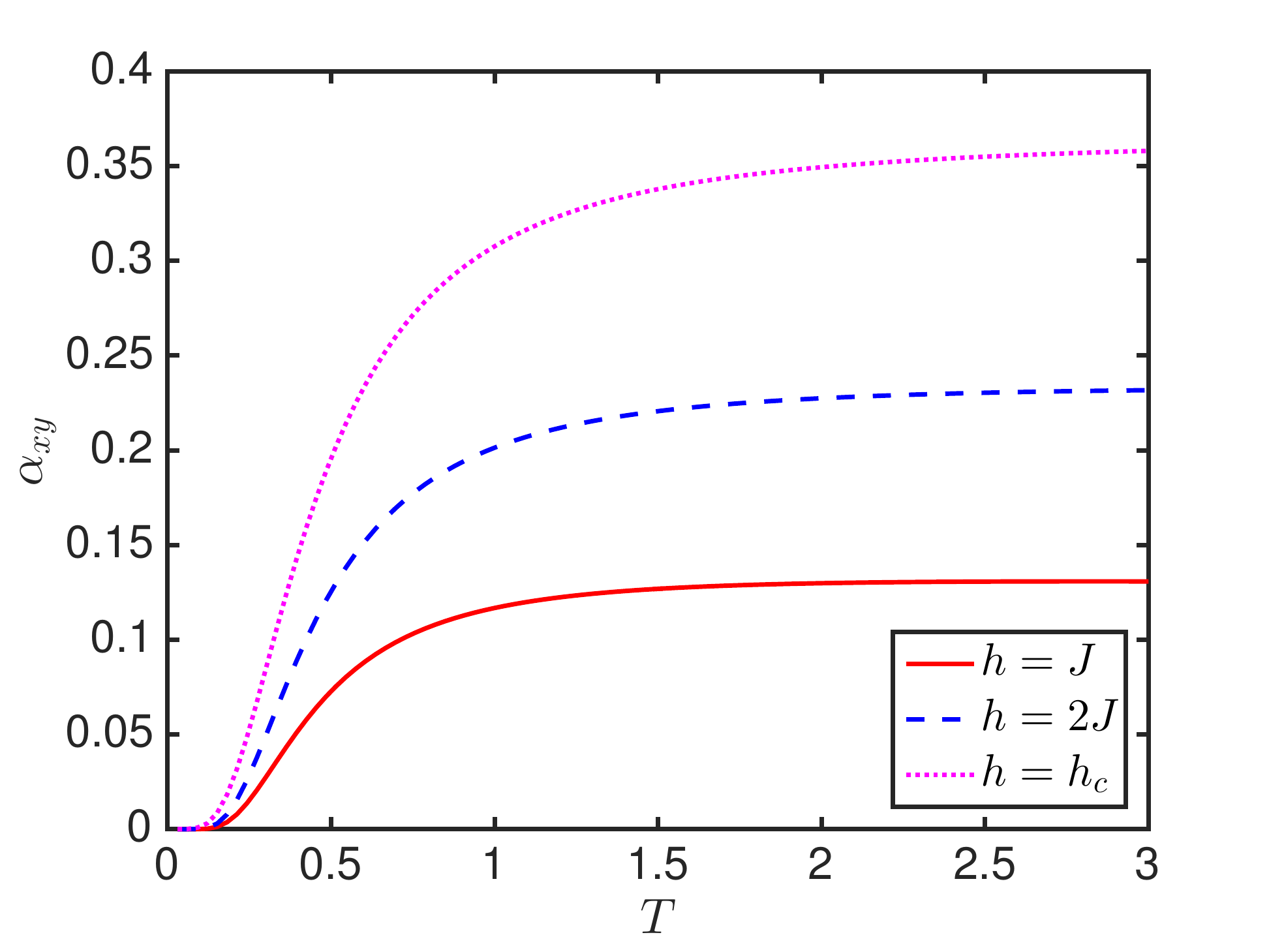}}
   \quad
   \subfigure[]{\includegraphics[width=.45\linewidth]{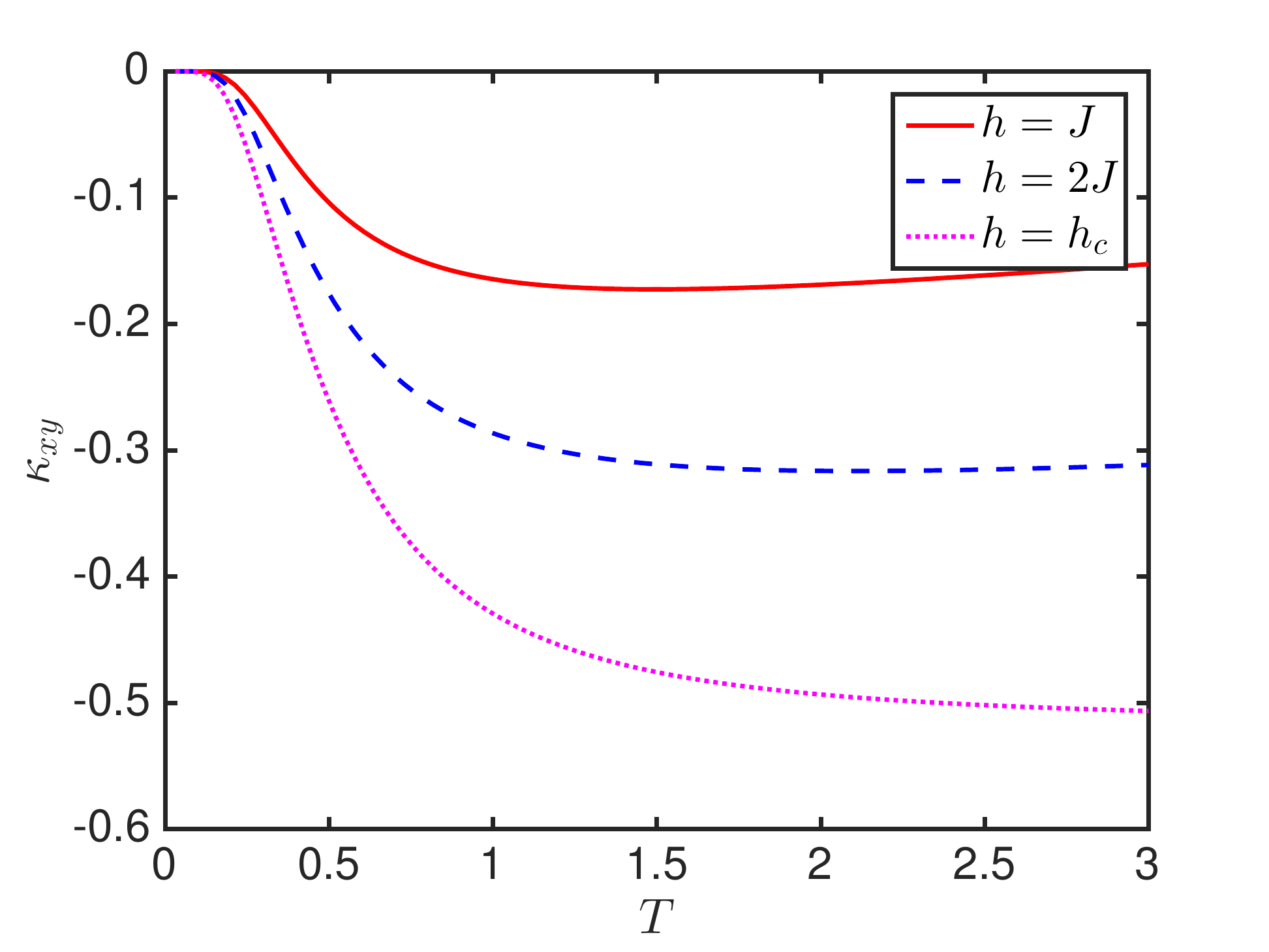}}
\caption{Color online. Spin Nernst conductivity  $(a)$ and Thermal Hall conductivity  $(b)$ of hardcore bosons in honeycomb lattice  with  $J=0.5;~D=0.25J$. We have set the natural constants to unity.}
\label{HCF}
\end{figure*}
Although the DM interaction does not contribute to the classical energy, the magnon excitations depend on it, thus a chiral superfluid is formed in the magnon bands. The two magnon bulk bands in chiral superfluid phase are shown in Fig.~\ref{band}.  At $\bold K (\bold K^\prime)$, $\gamma_{\bold k}=0$ and $m({\bold k})=m=\mp 3\sqrt{3}v_D$, hence  a gap  of magnitude $2|m|$ is generated in the magnon bulk bands. As mentioned above, the topological properties of this system are manifested once a gap opens at $\bold K (\bold K^\prime)$ in the magnon bulk bands.  From the bulk-edge correspondence, we expect  magnon edge states in the vicinity of the bulk gap. Indeed, we observe such edge states as shown in Fig.~\ref{band}. Due to the mapping from spin variables to bosonic operators and verse versa, the hardcore boson model \ref{hardcore} also exhibits bosonic edge states. 

An important consequence of gapped systems is the non-vanishing of the Berry curvature associated with the topology of the energy bands.  The Berry curvature associated with the gapped hard-core bosons is  given by 
\begin{equation}
\Omega_\lambda(\bold k)= -2 \textrm{Im}[\braket{\partial_{k_x}\psi_\lambda(\bold k)|\partial_{k_y}\psi_\lambda(\bold k)}],
\label{bb}
\end{equation}
where $\ket{\psi_\lambda(\bo)}$ are the positive eigenvectors.   It is more feasible to calculate the Berry curvature numerically. Hence, Eq.~\ref{bb} can be written as
\begin{align}
\Omega_\lambda(\bold k)=-\sum_{\lambda\neq \lambda^\prime}\frac{2\text{Im}[ \braket{\psi_{\bo\lambda}|v_x|\psi_{\bo\lambda^\prime}}\braket{\psi_{\bo\lambda^\prime}|v_y|\psi_{\bo\lambda}}]}{\lb\epsilon_{\bo\lambda}-\epsilon_{\bo\lambda^\prime}\rb^2},
\label{chern2}
\end{align}
where   $v_{i}=\partial \mathcal{H}_B(\bold k)/\partial k_{i}$ defines the velocity operators. Figure~\ref{berry} shows the Berry curvatures for the top and the bottom bands. It is obvious that the dominant contributions come from the states near $\bold K$ and $\bold K^\prime$.  In the low-energy limit, we expand the Hamiltonian near $\bold K (\bold K^\prime)$.  We obtain 

\begin{align}
{\mathcal{H}_B}(\bo)&= \sg_z[\bold{I}_\tau v_0 -\bar v_1(\xi \tau_x k_x+ \tau_y k_y)]\nonumber\\& -i\sg_y  \bar v_{2}(\xi \tau_x k_x+ \tau_y k_y)+\xi m{\bold I}_\sg\tau_z,
\label{hald}
\end{align}
where $\xi=\mp$ for states at $\bold K (\bold K^\prime)$ and $\bar v_{1,2}=3 v_{1,2}/2$. At  the fully polarized Mott states,  $\theta=0$ or $\theta=\pi$,  we have $\bar v_{1}=3v_0/2$ and  $\bar v_{2}=0$. Then, the model decouples  into two Bogoliubov Hamiltonians.  In this case, Eq.~\ref{hald} is similar to the Haldane model \cite{fdm}.  This is a direct consequence of the hard-core boson Hamiltonian \ref{hardcore}.

\subsection{Thermal and Spin Nernst conductivities}
In the preceding sections, we have explicitly demonstrated the nontrivial topological properties of the gapped honeycomb hardcore bosons. As mentioned in the Introduction, the magnon edge states carry a heat or spin current upon the application of a temperature gradient. This is, in fact, due  to  the nontrivial topological properties of the Berry curvature. These edge states lead to two important phenomena --- thermal Hall effect\cite{alex2} and spin Nernst effect. \cite{alex7} They are characterized by two conductivities given by\cite{alex2,alex7} 
\begin{align}
\alpha_{xy}&=\frac{k_B}{V}\sum_{\bo\lambda}c_1\lb n_\lambda\rb\Omega_\lambda(\bold k),\\
\kappa_{xy}&=-\frac{2k_B^2 T}{V}\sum_{\bo\lambda}c_2\lb n_\lambda\rb\Omega_\lambda(\bold k),
\end{align}
where $\alpha_{xy}$ is the spin Nernst conductivity and $\kappa_{xy}$ is the thermal Hall conductivity; $n_\lambda\equiv n_B[\epsilon_\lambda(\bold k)]=[e^{{\epsilon_\lambda(\bold k)}/k_BT}-1]^{-1}$ is the Bose function and the $c_i$ functions are $c_1(x)=(1+x)\ln(1+x)-x\ln x$; $c_2(x)=(1+x)\lb \ln \frac{1+x}{x}\rb^2-(\ln x)^2-2\textrm{Li}_2(-x),$ and $\text{Li}_n(x)$ is a polylogarithm.

 The plots of $\alpha_{xy}$ and $\kappa_{xy}$ as functions of temperature for  varying magnetic field are shown in Fig.~\ref{HCF}. The spin Nernst conductivity vanishes at zero temperature as there is no thermal excitations.  In stark contrast to the Heisenberg ferromagnet \cite{kkim}, it increases with increasing magnetic field as the system progresses from chiral superfluid phase to Mott phase and approaches a constant value at high temperature. The thermal Hall conductivity on the other hand is negative and never changes sign. At small magnetic field the system is in the chiral superfluid phase. As the magnetic field increases,  $\kappa_{xy}$ decreases as the system transits to the Mott insulator phase. 
 As mentioned above,  the hardcore boson model Eq.~\ref{hardcore} should exhibit the same  properties due to the mapping from spin variables to bosonic operators.

\section{Spinon formalism}

 The Schwinger boson mean-field theory\cite{alex10} provides another semiclassical approach to study the magnon  excitations in quantum magnets.  This method has been  employed to study the magnon bulk bands of SU(2) Heisenberg ferromagnets with DM interactions.\cite{kkim,alex5}  In this  representation, the spin operators are mapped to bosons using the following transformation \begin{align}
S_i^z =\frac{1}{2}\sum_{\sg}\sg c_{i,\sg}^\dg c_{i,\sg}; ~ S_i^+ = c_{i\up}^\dg c_{i\dw};~ S_i^- = c_{i\dw}^\dg c_{i\up},
\label{sch}
\end{align}
subject to the constraint $ \sum_{\sg}c_{i\sg}^\dg c_{i\sg}=2S=1,$ where $ \quad \sg = \up,\dw$ or $\pm$.   The Schwinger boson method is built upon the symmetry of the Hamiltonian. Since the XY model is not SU(2)-invariant, the bond operators are different.  They are given by\cite{alex11}
\begin{align}
&F_{ij}^\dg=\sum_{\sg}c_{i,\sg}^\dg c_{j,\sg}; \quad A_{ij}^\dg=\sum_{\sg\sg^\prime}\epsilon_{\sg\sg^\prime}c_{i,\sg}^\dg c_{j,\sg^\prime}^\dg;\\& X_{ij}^\dg =\sum_{\sg\sg^\prime}\tau_{\sg\sg^\prime}^x c_{i,\sg}^\dg c_{j,\sg^\prime}^\dg;\quad Z_{ij}^\dg=\sum_{\sg\sg^\prime}\tau_{\sg\sg^\prime}^z c_{i,\sg}^\dg c_{j,\sg^\prime},
\end{align}
where $\epsilon_{\up\dw}=-\epsilon_{\dw\up}=1$. The first two bond operators represent ferromagnetic and antiferromagnetic correlations and the last two represent easy axis and XY ferromagnetic correlations respectively. However, these bond operators are not independent. They are related by the  identities:
\begin{align}
&:F_{ij}^\dg F_{ij}: + A_{ij}^\dg A_{ij}=4S^2;\label{iden} \\& :Z_{ij}^\dg Z_{ij}: + X_{ij}^\dg X_{ij}=4S^2;
\label{iden1}
\end{align}
where $::$ represents normal ordering, {\it i.e.}, the creation operators placed to the left of the annihilation operators. 
The Schwinger boson Hamiltonian of Eq.~\ref{hh1} is given by
\begin{align}
&H=  -\frac{J}{2}\sum_{\la ij\ra}\big[:F_{ij}^\dg F_{ij}: +X_{ij}^\dg X_{ij}\big]\label{schh}\nonumber\\&- \frac{D}{2}\sum_{\la\la ij\ra\ra}\big[e^{i\nu_{ij}\phi}:F_{ij,\up}^\dg F_{ij,\dw:}+h.c. \big]\nonumber\\&+\sum_{i,\sg} (\lambda-\sg \frac{h}{2}) c_{i,\sg}^\dg c_{i,\sg},
\end{align}
where the constant terms have been dropped,  $F_{ij,\sg}^\dg=c_{i,\sg}^\dg c_{j,\sg}$, and $\lambda$ is a Lagrange multiplier. 
Next, we rotation the coordinate by $\pi/2$ about the $y$-axis, which changes the bond operator  $X^\dg$,
\begin{align}
X^\dg_{ij}=c^\dg_{i\up}c^\dg_{j\up}-c^\dg_{i\dw}c^\dg_{j\dw}.
\end{align}
The quartic operators in Eq.~\ref{schh} can be brought to quadratic form by performing mean field  Hartree-Fock decoupling
\begin{align}
:F_{ij}^\dg F_{ij}: &=Q^*F_{ij} + Q F_{ij}^\dg-|Q|^2,\\
X_{ij}^\dg X_{ij} &=P^*X_{ij} + P X_{ij}^\dg-|P|^2,\\
:F_{ij,\sigma}^\dg F_{ij,-\sg}: &=Q_\sg^*F_{ij,-\sg} + Q_{-\sg} F_{ij,\sg}^\dg-Q_\sg^*Q_{-\sg}.
\end{align}
Using the fact that $F_{ij,\sigma}=F_{ji,\sigma}^\dg$, we write $Q_\sg=Q_{1,\sg} +i\nu_{ij} Q_{2,\sg}$, where $Q_{12,\sg}$ are real variables. The resulting mean field Hamiltonian is given by
\begin{align}
&H=  -\frac{J}{2}\sum_{\la ij\ra,\sg}\big[Q c_{i,\sg}^\dg c_{j,\sg} +\sg P c_{i,\sg}^\dg c_{j,\sg}^\dg+h.c.\big]\nonumber\\&- \frac{D}{2}\sum_{\la\la ij\ra\ra,\sg}\big[e^{i\sg \nu_{ij}\phi} Q_{1,-\sg}c_{i,\sg}^\dg c_{j,\sg}+h.c. \big]\nonumber\\& -\frac{D}{2}\sum_{\la\la ij\ra\ra,\sg}\big[Q_{2,-\sg}\sg c_{i,\sg}^\dg c_{j,\sg}+h.c. \big]\nonumber\\&+\sum_{i,\sg} (\lambda-\sg \frac{h}{2}) c_{i,\sg}^\dg c_{i,\sg},
\label{sch}
\end{align}
The momentum space Hamiltonian can be written in the same form as the HP case with summation over spins and the Bogoliubov Hamiltonian is given by
 \begin{align}
\mathcal{H}_{B\sg}(\bo)&= m(\bold k){\bf I}_\sg \tau_z+ \sg_z[{\bold I}_\tau v_0(\bo)-\tilde v_{1}(\tau_+\gamma_\bo +h.c.)]\nonumber\\&- \tilde v_{2}i\sigma_y(\tau_+\gamma_\bo +h.c.),
\label{schw}
\end{align}
where $\tilde v_{1}=-z \sg J P/2, ~\tilde v_{2}=-z JQ/2$, $m(\bold k)=-\sg D Q_{1,-\sg}\rho_{\bold k}/2$,  $v_0(\bo)= (\lambda-\sg h/{2} -\sg D\tilde{\rho}_{\bold k}Q_{2,-\sg}/2)$, and $\tilde \rho_{\bf k}=2\sum_\mu\cos {\bf k}\cdot\bold{a}_\mu$. 
 To determine the $P's$ and the $Q's$ one needs to diagonalize Eq.~\ref{schw} using the Bogoliubov quasiparticles and solve a self-consistent equation. We do not perform this procedure here as it is not necessary since it gives quantitatively the same result as the conventional spin wave theory.  In fact,  experimental results  reported on the kagome ferromagnet \cite{alex6} suggest that the Holstein Primakoff transformation  is a better predictor than the Schwinger boson representation. The purpose of this section is to provide an alternative representation of the spin wave Hamiltonian. 
\section{Spinful Haldane-Hubbard model}
 The spinful Haldane-Hubbard model has recently garnered much attention.\cite{pa,pa2,pa3} The Hamiltonian is governed by
\begin{align}
H&= -t\sum_{\langle ij\rangle\sigma}( b^\dg_{i\sigma} b_{j\sigma} +  h.c.) - t^\prime \sum_{\langle\langle ij\rangle\rangle\sigma}(e^{i\nu_{ij}\phi}b_{i\sigma}^\dg b_{j\sigma}+ h.c.)\\&\nonumber+\frac{U}{2}\sum_{i\sigma\sigma^\prime} n_{i\sigma}n_{i\sigma^\prime}.
\end{align}
We restrict the operators to obey only the boson commutation relation $[b_{i\sg}, b_{j\sg^\prime}^\dg]=\delta_{ij}\delta_{\sg\sg^\prime}$. The phase diagram of this model for fermionic systems have been discussed in Ref.~\cite{pa,pa2,pa3}. For bosons, it has not been studied extensively in the topological context.  Considering only the flux enclosed by the unit cell in the big triangular plaquettes of the NNN sites, in the strong coupling limit $U/t\gg 1$; $U/t^\prime\gg 1$, the spinful Haldane-Hubbard model maps to \cite{pa}\begin{align}
H&=-J\sum_{\la ij\ra}{\bf S}_{i}\cdot{\bf S}_{j}-J^\prime\sum_{\la \la ij\ra\ra}{\bf S}_{i}\cdot{\bf S}_{j}+D\sum_{\Delta,\nabla} \chi_{\Delta,\nabla}.
\label{h}
\end{align}
The coupling constants are given by $J=\frac{4t^2}{U}$, $J^\prime=\frac{4t^{\prime 2}}{U}$, $D=\frac{24t^{\prime 3}}{U^2}$, and we have set $\phi=\pi/2$. The last term sums over the triangular plaquettes on the NNN sites. It corresponds exactly to the DM interaction as discussed above. In this model, the interactions are ferromagnetic.   Quite remarkably, Eq.~\ref{h} is exactly the model studied in Ref.~\cite{sol,sol1, kkim} These results show a strong relationship  between bosonic systems and quantum spin systems.


\section{Conclusion}
The main result of this  paper is that the topological properties of hardcore bosons and that of magnons in quantum magnets are not independent. This is because topological effects manifest in the excitations of the corresponding system and the  hardcore bosons correspond to quantum magnetic systems.   In the spin language, we have shown that the Haldane-hardcore bosons exhibit interesting nontrivial topological properties. This suggests that the bosonic excitations of hardcore bosons (which have to be magnons) possess the same topological properties. Although QMC is not applicable in the present model, other  numerical schemes should provide further insights into these topological effects. We believe that these nontrivial topological  effects should also manifest in frustrated  systems by including a next-nearest-neighbour antiferromagnetic interaction.   We also showed that in the strong coupling limit of the spinful Haldane-Hubbard model on the honeycomb lattice, the system maps to Heisenberg ferromagnet whose topological magnon excitations have been studied recently. 

Though the noninteracting hardcore boson model captures topologically  nontrivial properties, in most cases of physical interest interactions are not negligible. The simplest interaction   is the repulsive NN interaction given by \cite{var1}  \begin{align}
&H_{int}= V\sum_{\langle ij\rangle}n_i n_j \label{hamm},
\end{align}
where $V>0$. On the honeycomb lattice, such interaction does not introduce any frustration in the system. It is easy to show that in the spin variables,  $H_{int}=J_z\sum_{\langle ij\rangle}S_i^z S_j^z$ does not contribute to the gap opening at $\bold K (\bold K^\prime)$, hence its effects are negligible in the topological considerations. Furthermore, we claim that these topological properties should also be observed in frustrated systems \cite{arg}.  For the XY model on the honeycomb lattice, frustration is induced by an antiferromagnetic NNN coupling \cite{var}.  The system also possesses ordered states in the phase diagram, and the excitations of these ordered states should exhibit  similar topological properties when a nontrivial gap is introduced. However, the explicit analysis of this frustrated  model is beyond the purview of this paper. The correspondence between magnetic spins systems and hardcore bosonic systems suggests experimental procedures in cold atoms on the honeycomb optical lattices to search for these interesting topological properties in bosonic and magnetic systems. 
\section*{Acknowledgments}
The author would like to thank African Institute for Mathematical Sciences (AIMS). Research at Perimeter Institute is supported by the Government of Canada through Industry Canada and by the Province of Ontario through the Ministry of Research
and Innovation.

\end{document}